\documentclass[12pt,a4paper]{article}
\usepackage{graphicx}
\usepackage{dcolumn}
\usepackage{bm}
\usepackage{amsmath}
\usepackage{amsfonts}%
\usepackage{amssymb}
\usepackage[mathscr]{eucal}
\usepackage{cite}
\renewcommand{\baselinestretch}{1.5}
\begin{document}

\begin{center}
\vskip 48 pt
{\Large \bf An analytical study of electronic properties of ABC-stacking multilayer graphene }
\vskip 12 pt
Cheng-Peng Chang $^{*}$\footnotetext{Corresponding author. Tel/Fax: +886-62-545329\\
     E-mail address:  t00252@mail.tut.edu.tw (C. P. Chang)}\\
{\small Center for General Education,  Tainan  University of Technology, Tainan 710, Taiwan}\\
\end{center}
\renewcommand{\baselinestretch}{2}
\vskip 12 pt
\begin{abstract}
We present an analytical model to study the electronic properties, including full band structure, low energy dispersions around the Dirac point and density of states of the ABC-stacking $N$-layer graphene (ABCNLG).  An ABCNLG can be simulated by a linear atomic chain with $2N$ atoms. With only nearest-neighbor inter- and intra-layer hopping integrals taken into account, the Hamiltonian representation is a complex $2N \times 2N$ tridiagonal matrix $H_0$. Through a unitary transformation, we can reduce the $2N \times 2N$ Hamiltonian matrix into two real $N \times N$ tridiagonal matrices $\mathbb{H}_{s}$ and $\mathbb{H}_{a}$, i. e., $H_0=\mathbb{H}_{s} \oplus \mathbb{H}_{a} $.  What's more, the two matrices satisfy the relation $\mathbb{H}_{a}=-\mathbb{H}_{s}$.  As a result, energy spectrum associated with $\mathbb{H}_{s}$  and $\mathbb{H}_{s}$ have the relation $\lambda_{a}=-\lambda_{s}$. Such a characteristic is reflected on the energy dispersions and density of states. Our model can be applied to explore the basic properties of linear chain model and the eigenvalue problem of the tridiagonal matrices.
\end{abstract}
\pagebreak
\renewcommand{\baselinestretch}{0.8}

\section {Introduction}
Graphene, a pristine two-dimensional (2D) material,  is isolated by the  exfoliation method\cite{Novoselov1}, and it also  offers a display place to exhibit fundamental properties of 2D system. Due to the specular geometry structure, graphene shows many interesting electronic properties, e. g., low-lying linear energy bands, electron-hole symmetry, high room-temperature mobility, high in-plane thermal conductivity  Klein tunneling,  and anomalous quantum Hall
effect\cite{Novoselov2,Zhang2,Gusynin,McCann,Peres,Abergela,Basov2014,Castro}. Multilayer graphene, one of the carbon allotropes, is the pile of several graphene layers, bound  together  by the van der Waals interactions, along the stacking direction.  The low-energy electronic structures are strongly related to the stacking types  and the number of
layers\cite{Castro,Latil,Lu2,LU2007,Graf,Nilsson,Koshino2008,Reina,Avetisyan,Craciun,Chun-Hung-Lui, Chang-2012,Chang-2012-1,McCann2013, Rozhkova2016}. The usually studied stacking sequences of graphene multilayers are the Bernal-stacking (AB-stacking), simple hexagonal stacking (AA-stacking), and  rhombohedral stacking (ABC-stacking).
Bernal-stacking (AB-stacking) bilayer graphene  has attracted intense interest because its bandgap can be controlled by  applying  to an external electric field\cite{Latil,Chang-2012,Chang-2012-1,McCann2013, Rozhkova2016}.  With the  controllable  bandgap,   Bernal-stacking  bilayer graphene shows  great potential as a new material for opto-electronic devices.  In the absence of the external field,  Bernal bilayer graphene is  semimetal because of the tiny touch  between valence and conduction bands.   Two groups of the parabolic band around the Dirac points are presented in  Bernal bilayer graphene\cite{Latil,Chang-2012}.  The electronic properties of AB-stacking multilayer graphenes also inspire a large number of studies owing to possible applications. Study results exhibit that the AB-stacking $N$-layer graphenes with even $N$ layers (with odd $N$ layers) are to be equivalent to the superposition of $N/2$ bilayer graphenes ($ \frac{N-1}{2}$ bilayer graphene and one graphene-like monolayer) \cite{Koshino2008,Chang-2012-1 }.


Before the experimental realization of the AA-stacking graphite, the AA-stacking bilayer graphene, due to its simple geometrical structure, is usually  utilized as a theoretical model to demonstrate the low-energy electronic properties, two pairs of nearly linear bands, which are distinguishable from those of the AB-stacking bilayer graphene because of the different stacking types. Recently, the fabrication of AA-stacking graphite\cite{JKLee2008}  and the AA-stacking multilayer graphenes\cite{Borysiuk2011}  renew interest in  the fundamental properties of AA-stacking bilayer and multilayer graphenes, e.g. infrared and Raman spectra, Landau-level energies, transport, plasma excitations,  magneto-absorption spectra and dynamical conductivity\cite{Liu-2009,YuehuaXu,Chiu,
Chang2011,Tabert2012,Ming-Fa-Lin2012,Chang2013,Chang2015}. The study results exhibit that the electronic properties of the AA-stacking $N$-layer graphene can be treated as the superposition of $N$  independent graphene-like monolayers\cite{Chang2013,Chang2015}.

The electronic properties of  an ABCNLG, e. g. , the flat band, bulk band, bulk band gap, electron velocity, and density-of-state, are strongly dependent  on stacking order and modified by the application of a vertical electric field\cite{Koshino2010,Zhang2010,Yuan2011,Koshino2013,
Yi-Ping-Lin2015,Ching-Hong2016}.
They are clearly  distinct from  those of AA- or AB-stacking multilayer graphenes. Most studies focus on the low-energy electronic structures of ABC-stacking trilayer graphene. A non-perturbative effective Hamiltonian closed in the bulk subspace is proposed to  explore the bulk subbands  for arbitrary  the layer number $N$ \cite{Ching-Hong2016}.  Recently,  experimental approval of the extended flat bands and the gapped subbands in ABC-stacking multilayer graphene has been reported\cite{Henni2016,Bao2017}.
The  above-mentioned experimental works trigger us to revisit the electronic  properties of ABCNLG. We present an analytical model to effectively and efficiently   study  full band structure, low energy dispersions around the Dirac point and density of states.

\section{Theory and Model}

The geometrical structure of an ABCNLG is shown in Fig. 1(a). Each graphene layer is a one-atom-thickness layer, in which carbon atoms are precisely packed in a hexagonal lattice. The lattice contains two sublattices $A$ and $B$,  represented by white and black circles, respectively.  A primitive cell contains two atoms and the nearest carbon-carbon distance is $b= 1.42{\rm \AA}$. Within ABCNLG, half of the atoms are directly below atoms in the adjacent sheet and directly above hexagonal ring centers and the other half of the atoms are directly above atoms and directly below hexagonal ring centers\cite{Charlier3}. A primitive cell of an ABCNLG has ${2 N}$ carbon atoms, denoted as, $A_1, B_1, A_2, B_2, \cdots,  A_{l}, B_{l}, \cdots, A_{N}, B_{N}$. The first Brillouin zone is also shown in Fig. 1(a). The atom-atom interactions, shown in Fig. 1(a), are as follows. $\beta_0$ represents the interaction between atom $A$ and $B$ on the same graphene layer. The interlayer interaction between the atom $A$ and  $B$ from the two neighboring layers is $\beta_1$ while the two atoms have the same $(x,y)$ coordinate. The distance between the two nearest neighboring sheets is $c=3.35~{\rm \AA}$. The values of $\beta_0$ and $\beta_1$ are \cite{Charlier3} $\beta_0 =3.16$ eV and $\beta_1 =0.36$ eV.

The Hamiltonian representation $H_0$, expanded in the set of 2D Bloch functions,
$(|A_1\rangle$,  $|B_1\rangle$,   $|A_2\rangle$, $|B_2\rangle$, $\cdots \cdots$,    $|A_{N-1}\rangle, |B_{N-1}\rangle$, $|A_{N}  \rangle, |B_{N}  \rangle)$, is a $2N \times 2N$ matrix and reads
\begin{eqnarray}
H_0=
\left(
\begin{array}{cccccccccccc}
0& \beta_0 {\bf f_k}  &0 &0 &0 &   \cdots&\cdots& 0  &0 & 0  &0 & 0 \\
\beta_0  {\bf f^*_k} &0& \beta_1 &0 &0 &\cdots &\cdots&0&0& 0&0&0 \\
0 & \beta_1& 0 &  \beta_0{\bf f_k} &0&\cdots &\cdots& 0&0& 0&0&0 \\
0 &0   &\beta_0 {\bf f^*_k}& 0 & \beta_1 &\cdots   &\cdots  &0&0& 0&0&0 \\0 &0 &0 &\beta_1  &\ddots &\ddots &\cdots&0& 0&0&0 &0 \\
\vdots &\vdots&\vdots &\vdots& \ddots &\ddots &\ddots &0&\vdots &\vdots&\vdots &\vdots \\
\vdots &\vdots&\vdots &\vdots&0 &\ddots &\ddots &\ddots&\vdots &\vdots&\vdots &\vdots \\
0 &0&0  &0  &0 &\cdots &\ddots &0&\beta_1& 0&0&0 \\
0 &0&0 &0 & 0 &\cdots &\cdots &\beta_1&0& \beta_0{\bf f_k}&0&0 \\
0 &0&0 &0 & 0 &\cdots &\cdots &0&\beta_0{\bf f^*_k}& 0&\beta_1&0 \\
0&0&0& 0&0&\cdots  &\cdots &0 &0&\beta_1 & 0 & \beta_0{\bf f_k} \\
0&0& 0&0& 0  & \cdots & \cdots& 0 &0& 0 &  \beta_0{\bf f^*_k}& 0 \cr
\end{array}
\right)_{2N \times 2N},
\label{HM-1}
\end{eqnarray}
where $ {\bf f_k}=\sum ^3_{j=1} {\rm exp} (i {\bf k}\cdot {\bf b}_j)= f e^{i \theta}$ and $\theta= tan^{-1}(\frac{Im[~\bf f_k~]}{Re[~\bf f_k~]})$; ${\bf b}_j$ represents the three nearest neighbors on the same graphene plane and ${\bf k}$ is the in-plane wave vector. Obviously, $H_0$ is a tridiagonal matrix  with  complex  elements resulting from the complex structure factor $ {\bf f_k}$. It is more efficient to diagonalize a real matrix for eigenvalues than a complex one. Such a complex tridiagonal matrix (Eq. (\ref{HM-1})) can be easily transformed into a real symmetrical matrix through a unitary transformation. By adopting the new 2D Bloch functions
($ \phi_1= |A_1\rangle,~
   \phi_2 = e^{i \theta}  |B_1\rangle,~
  \phi_3 = e^{i \theta}  |A_2\rangle,~
  \phi_4 = e^{i 2 \theta}|B_2\rangle,$~
   $\cdots \cdots,
      \phi_{2N-3}=e^{i(N-2)\theta} |A_{N-2}\rangle$,~
    $\phi_{2N-2}=e^{i(N-1)\theta} |B_{N-l}\rangle,$~
    $\phi_{2N-1}=e^{i(N-1)\theta} |A_{N}  \rangle$,~
    $\phi_{2N}    =e^{i N   \theta} |B_{N}  \rangle) $,
  we  calculate the Hamiltonian matrix element  ${\mathcal H} (i,j)=\langle \phi_i| H_0|\phi_j\rangle$ and the transferred Hamiltonian representation is obtained as follows
\begin{eqnarray}
{\mathcal H}=
\left(
\begin{array}{cccccccccccc}
0& \beta_0 f  &0 &0 &0 &   \cdots&\cdots& 0  &0 & 0  &0 & 0 \\
\beta_0   f &0& \beta_1 &0 &0 &\cdots &\cdots&0&0& 0&0&0 \\
0 & \beta_1& 0 &  \beta_0 f &0&\cdots &\cdots& 0&0& 0&0&0 \\
0 &0   &\beta_0  f & 0 & \beta_1 &\cdots   &\cdots  &0&0& 0&0&0 \\
0 &0 &0 &\beta_1  &\ddots &\ddots &\cdots&0& 0&0&0 &0 \\
\vdots &\vdots&\vdots &\vdots& \ddots &\ddots &\ddots &0&\vdots &\vdots&\vdots &\vdots \\
\vdots &\vdots&\vdots &\vdots&0 &\ddots &\ddots &\ddots&\vdots &\vdots&\vdots &\vdots \\
0 &0&0  &0  &0 &\cdots &\ddots &0&\beta_1& 0&0&0 \\
0 &0&0 &0 & 0 &\cdots &\cdots &\beta_1&0& \beta_0 f &0&0 \\
0 &0&0 &0 & 0 &\cdots &\cdots &0&\beta_0 f& 0&\beta_1&0 \\
0&0&0& 0&0&\cdots  &\cdots &0 &0&\beta_1 & 0 & \beta_0 f \\
0&0& 0&0& 0  & \cdots & \cdots& 0 &0& 0 &  \beta_0 f& 0 \cr
\end{array}
\right)_{2N \times 2N}.
\label{HM-2}
\end{eqnarray}
The Hamiltonian matrix~of  an ABCNLG is now transformed into a real and symmetrical matrix.

According to Eq. (\ref{HM-2}), an ABCNLG can be modeled as a linear atomic chain  of carbon atoms, as shown in Figs. (1b) and (1c). In order to reduce the dimension of Hamiltonian matrix, we further construct the  symmetrized basis functions:
$|\psi^s_1$,
$\psi^s_2, \cdots$,
$\psi^s_i, \cdots$,
$\psi^s_{N-1}$,
$\psi^s_N$,
$\psi^a_1$,
$\psi^a_2, \cdots$,
$\psi^a_i, \cdots$,
$\psi^a_{N-1}$,
$\psi^a_N\rangle $.
Based on the mirror symmetry, or inversion symmetry,  of linear atomic chain, they are divided into two groups, symmetrical and anti-symmetrical groups, and organized as follows:
$$
\begin{cases}
\psi^s_1= (\phi_1 +\phi_{2N})/\sqrt{2}\cr
\vdots\cr
\psi^s_j= (\phi_j +\phi_{2N+1-j})/\sqrt{2}\cr
\vdots\cr
\psi^s_N=(\phi_N +\phi_{N+1})/\sqrt{2}\cr
\psi^a_1=(\phi_1 -\phi_{2N})/\sqrt{2}\cr
\vdots \cr
\psi^a_j=(\phi_j -\phi_{2N+1-j})/\sqrt{2}\cr
\vdots \cr
\psi^a_N=(\phi_N -\phi_{N+1})/\sqrt{2}\\
\end{cases}
$$
After some manipulation, the $2N \times 2N$ Hamiltonian matrix $H_0$ is decomposed into two $N \times N$ diagonal-block matrices, i. e.,  the reduced  Hamiltonian matrix reads
\begin{eqnarray}
\mathbb{H}_{red}=
\left(
\begin{array}{cc}
     \mathbb{H}_{s}   &  0  \\
     0   &  \mathbb{H}_{a}
\end{array}
\right).
\label{HMreduce}
\end{eqnarray}
It is noted that  the two diagonal-block matrices satisfy the relation $\mathbb{H}_{s}(i,j)= - \mathbb{H}_{a}(i,j)$.
This is to say, the reduced Hamiltonian representation has the characteristics, $\mathbb{H}_{red}=  \mathbb{H}_{s}  \oplus \mathbb{H}_{a}$ and $\mathbb{H}_{s}= - \mathbb{H}_{a}$.

 The representation actually $\mathbb{H}_{s}$ depends on the layer number $N$. When $N=2 m+1$ is odd, the representation of the block matrix $\mathbb{H}_{s}$ is
\begin{eqnarray}
 \mathbb{H}_{s}=
\left(
\begin{array}{ccccccccc}
0& \beta_0 f  &0 &0 &0  &\cdots&\cdots       &0 &0  \\
\beta_0 f &0& \beta_1 &0 &0 &\cdots &\cdots   &0 &0\\
0 & \beta_1& 0 &  \beta_0 f &0&\cdots &\cdots&  0  &  0\\
0 &0 &\beta_0 f & 0 & \beta_1 &\cdots   &\cdots  &0&0 \\
0 &0 &0 &\beta_1  &\ddots &\ddots &\cdots     &0& 0    \\
\vdots &\vdots&\vdots &\vdots& \ddots &\ddots &\ddots &\vdots &\vdots  \\
\vdots &\vdots&\vdots &\vdots& \ddots &\ddots &\ddots &\beta_0f &\vdots  \\
0&0  &0  &0 &\cdots &\ddots &\beta_0f & 0 &\beta_1\\
0&0  &0 &0 & 0 &\cdots &\cdots &\beta_1 & \beta_0f  \\
\end{array}
\right)_{N\times N}.
\label{HMatrix}
\end{eqnarray}
$ \mathbb{H}_{s}$ is a real tridiagonal matrix with a non-zero element occurring at the corner $ \mathbb{H}_{s}(N,N)= \beta_0f  $.
As is shown in Fig. 1(b), the ABC-stacking trilayer graphene is described by an atomic chain with $2N=6$ atoms. The decomposition of $\mathcal H$ (Eq.(\ref{HM-2})) into $\mathbb{H}_{s}$   and $\mathbb{H}_{a}$ (Eq.(\ref{HMreduce})) is equivalent to cutting a long atomic chain into two short atomic chains and each chain is made up of $N=3$ atoms.  One surface atom of the short chain has the site-energy $\beta_0f$ (or $-\beta_0f$). On the other hand,
\begin{eqnarray}
 \mathbb{H}_{s}=
\left(
\begin{array}{cccccccc}
0& \beta_0 f  &0 &0 &0  &\cdots&\cdots       &0   \\
\beta_0 f &0& \beta_1 &0 &0 &\cdots &\cdots   &0 \\
0 & \beta_1& 0 &  \beta_0 f &0&\cdots &\cdots&  0  \\
0 &0 &\beta_0 f & 0 & \beta_1 &\cdots   &\cdots  &0 \\
0 &0 &0 &\beta_1  &\ddots &\ddots &\cdots     &0    \\
\vdots &\vdots&\vdots &\vdots& \ddots &\ddots &\ddots &\vdots \\
\vdots &\vdots&\vdots &\vdots& \ddots &\ddots &\ddots &\beta_0f \\
0&0  &0  &0 &\cdots &\ddots &\beta_0f &\beta_1\\
\end{array}
\right)_{N\times N},
\label{HMatrix}
\end{eqnarray}
when the layer number $N (=2 m)$ is even. For instance, the ABC-stacking quad-layer graphene is simulated by an  8-atom  chain, as is shown in Fig. 1(c). The reduced Hamiltonian matrix  $\mathbb{H}_{s}$ ($\mathbb{H}_{a}$) is modeled by a 4-atom chain (the right panel of Fig. 1(c). Each 4-atom chain has two asymmetrical surface atoms. The renormalized surface atom has site-energy $\beta_1$ ($-\beta_1$).

The eigen-equation of ABCNLG (${H}_{0}|\textbf{u}_0\rangle= \lambda |\textbf{u}_0\rangle$) is now decomposed into two eigen-equations, which are
\begin{eqnarray}
\mathbb{H}_{\xi}|\textbf{\emph{u}}_{\xi}\rangle= \lambda_{\xi}|\textbf{\emph{u}}_{\xi}\rangle,
\label{Ei-EQ}
\end{eqnarray}
where $\xi= s$ or $a$ denotes the symmetrical or anti-symmetrical state.  The eigenenergy spectrum $\lambda_{a}=-\lambda_{s}$ because of  $\mathbb{H}_{a}= - \mathbb{H}_{s}$. Moreover, $\mathbb{H}_{a}$ and $\mathbb{H}_{s}$ share the same eigenvector $|\textbf{\emph{u}}_{\xi}\rangle$, i. e., $|\textbf{\emph{u}}_{s}\rangle=|\textbf{\emph{u}}_{a}\rangle$.
Once the eigenenergy spectrum $ \lambda_{s}$ of $\mathbb{H}_{s}$ is acquired,  the eigenenergy spectrum $ \lambda_{a}$ is thus obtained through the relation $\lambda_{a}=-\lambda_{s}$.

\section {Energy spectra of ABCNLGs}
\subsection {Energy spectrum of $N=3$ ABCNLG}
The reduction of the size of Hamiltonian matrix allows us easily and efficiently to acquire the eigenenergies of ABCNLG. The  analytical form of eigenenergy of $N=3$ ABCNLG is obtained by solving the secular equation $| \mathbb{H}_{s}- \lambda \mathbb{I} |=0$, where $\mathbb{I}$ is the identity matrix and $\lambda$ is eigenvalues.
The secular equation associated with the trilayer graphene is
\begin{eqnarray}
| \mathbb{H}_{s} - \lambda  \mathbb{I}|= det
\left|
\begin{array}{ccc}
-\lambda& \beta_0 f  &0 \\
\beta_0 f &-\lambda& \beta_1 \\
0 & \beta_1&   \beta_0 f-\lambda  \\
\end{array}
\right|_{3\times 3}
=0.
\label{det-3}
\end{eqnarray}
The secular equation is a cubic polynomial
$\lambda^3 + r \lambda^2 + s \lambda + t=0,$ where
the coefficients are $r=- \beta_0 f$,  $s= -(\beta^2_0 f^2 +\beta^2_1 )$ and $t= \beta^3_0 f^3$. The roots of secular equation are the eigenenergy spectrum $\lambda^{(N)}_j$ ($j=1,2,3$). The close form of $\lambda^{(3)}_j$  is
\begin{eqnarray}
\begin{cases}
\lambda^{(3)}_1= -r/3 +\big[P- \sqrt{Q^3 +P^2}]^{1/3} +[P + \sqrt{Q^3 +P^2}]^{1/3},\\
\lambda^{(3)}_2= -r/3 + e^{\pi/3}[P- \sqrt{Q^3 +P^2}]^{1/3} - e^{-\pi/3}[P + \sqrt{Q^3 +P^2}]^{1/3},\\
\lambda^{(3)}_2= -r/3 + e^{-\pi/3}[P- \sqrt{Q^3 +P^2}]^{1/3} - e^{\pi/3}[P + \sqrt{Q^3 +P^2}]^{1/3},\\
\end{cases}
\label{eigen-3}
\end{eqnarray}
where $\displaystyle P=\frac{-r^3}{27} +\frac{st}{6} -\frac{t}{2}$ and $\displaystyle Q=\frac{s}{3}-\frac{r^2}{9}$.

\subsection {Energy spectrum of $N=4$ ABCNLG}

 The secular equation realted to ABCNLG with $N=4$ is
\begin{eqnarray}
| \mathbb{H}_{s} - \lambda  \mathbb{I}|= det
\left|
\begin{array}{cccc}
-\lambda & \beta_0 f  &0   &0   \\
\beta_0 f &-\lambda  & \beta_1      &0    \\
0 & \beta_1 & -\lambda &  \beta_0 f  \\
0 & 0 &\beta_0 f  & \beta_1-\lambda  \\
\end{array}
\right|_{4\times 4}=0.
\label{det-4}
\end{eqnarray}
It is a fourth order polynomial $\lambda^4 - \beta_1 \lambda^3 -(\beta^2_1 + 2\beta^2_0 f^2 ) \lambda^2 + \beta_1 (\beta^2_1 + \beta^2_0 f^2 ) \lambda + \beta^4_0 f^4 =0$. The roots $\lambda^{(4)}_j$ of the secular equation are the energy spectrum.

\subsection {The calculated energy spectrum}

Energy spectrum of  ABC-stacking $N$-layer graphene can be acquired by the exact diagonalization using numerical l  ibrary or by solving the roots of the associated secular equation. First, the energy dispersions of ABC-stacking trilayer graphene, as is shown in Fig. 2(a), are used to check the correction of our model. The red curves are the energy dispersions of the symmetrical Hamiltonian $\mathbb{H}_{s}$. They are the eigenvalues obtained by the numerical diagonalization  method. We further apply the relation $\lambda_{a}=-\lambda_{s}$  to gain the  energy dispersions of the asymmetrical Hamiltonian $\mathbb{H}_{a}$, the green curves. The dashed curves are calculated  by employing  Eq.(\ref{eigen-3}), the close form of the eigenenergy. The dashed curves are completely identical to the red curves.

As is shown in Fig. 2(b), the energy dispersions of ABC-stacking quad-layer ($N=4$) graphene are obtained by the diagonalization of  matrices, $H_0$, ${\mathcal H}$ and $\mathbb{H}_{s}$ (Eq. (\ref{HM-1}),  Eq.(\ref{HM-2}) and Eq.(\ref{HMreduce})). The entire overlap of three sets of energy dispersions exhibits that the reduced Hamiltonian matrix  $\mathbb{H}_{s}$ can efficiently offer the eigenenergy spectrum.

An ABCNLG illustrates $N$ branches of the energy dispersion.  Each branch is closely similar to the energy dispersions of a monolayer graphene, such as the maximum of the energy dispersion occurring at the point $\bf \rm \Gamma$,  the discontinuity presented near the saddle point $\bf \rm M$ and the nearly linear bands appearing around the  $\bf \rm K$ point.

\subsection{Low-energy spectrum around the Dirac point}
Our model can clearly reveal the characteristic of the low-energy electronic structures. Figures 3(a)-3(h) present the low-energy bands of ABCNLG with the layer number $N$. There are $N$ red  (green) curves, which are the energy dispersions of Hamiltonian $\mathbb{H}_{s}$ ($\mathbb{H}_{a}$). The red  curves are symmetrical to the green ones about the Fermi energy $E_F=0$. The electronic structures of an ABCNLG  show linear bands, flat bands and Mexican-hat bands. The flat bands, existing near the $E_F$, originate in the surface state.  The extension of the flat bands increases with the increase of the number of layers. The remainder bands are the $2N-2$ bulk bands. The highest conduction bands  and the lowest valence bands are nearly linear bands.  An AA-stacking $N$-layer graphene presents $N$ pairs of linear bands. An AB-stacking multilayer graphene illustrates parabolic bands and linear bands.

 The low-energy bands of an ABCNLG also depend on whether the layer number $N$ is odd or even. The red curves shows one surface state, $\frac{N-1}{2}$ conduction bands and  $\frac{N-1}{2}$ valence bulk bands when $N=2m+1$ is odd. At the Dirac point $\bf K$, $f_{\bf  k}=0$ and the eigenvalue of $\mathbb{H}_{s}$ are $\lambda_j= 0,~\pm\beta_1,~\pm\beta_1,\cdots,\pm\beta_1$. On the other hand, there are  one surface state, $\frac{N}{2}$ conduction bands and  $\frac{N}{2}-1$ valence bulk bands when $N=2m$ is even. At the Dirac point $\bf K$, the eigenvalue of $\mathbb{H}_{s}$
is $\lambda_j=0,\beta_1,\pm\beta_1,\pm\beta_1,\cdots,\pm\beta_1$.

The properties of the flat bands are  investigated.  The flat bands is closely located at $\lambda\sim 0$. By setting $\lambda\sim 0$ in the secular equation and neglecting the high order terms of $\lambda$, the secular equation related to  ABC-stacking trilayer graphene is $ -(\beta^2_0 f^2 +\beta^2_1 )\lambda +\beta^3_0 f^3=0$ and, thus, the eigenenergy of the flat band is obtained  $\lambda=\frac{\beta^3_0 f^3}{\beta^2_0 f^2 +\beta^2_1}\approx \frac{\beta^3_0 f^3}{\beta^2_1}$  because of $\beta_0 f \ll \beta_1$ in the vicinity of the point $\rm \bf K$. The secular equation of ABC-stacking quad-layer graphene is approximated as $\beta_1 (\beta^2_1 + \beta^2_0 f^2 ) \lambda + \beta^4_0 f^4 =0$  and the eigenenergy of the flat band  reads  $\lambda=- \frac{\beta^4_0 f^4}{ \beta_1 (\beta^2_1 + \beta^2_0 f^2 )}\approx  - \frac{\beta^4_0 f^4}{ \beta_1^3} $.

The wave function associated with the flat band of an ABCNLG is calculated. The eigenenergy of the flat band of ABC-stacking trilayer graphene is $\lambda=\frac{\beta^3_0 f^3}{\beta^2_1}$. After inserting $\lambda$ into eigen equation (Eq.(\ref{Ei-EQ})), we acquire the eigen equation as follows
\begin{eqnarray}
\left[
\begin{array}{ccc}
 0 & \beta_0 f  &0 \\
\beta_0 f & 0 & \beta_1 \\
0 & \beta_1&   \beta_0  \\
\end{array}
\right]
\left[
\begin{array}{c}
u_1 \\
u_2 \\
u_3 \\
\end{array}
\right]
=\frac{\beta^3_0 f^3}{\beta^2_1}
\left[
\begin{array}{c}
u_1 \\
u_2 \\
u_3 \\
\end{array}
\right].
\label{eigen eq-3}
\end{eqnarray}
The transverse of the eigen state $|\textbf{\emph{u}}_{\xi}\rangle^T $ is $(u_1 ,u_2 ,u_3 )=(1,\frac{\beta^2_0 f^2}{\beta^2_1},  \frac{\beta^5_0 f^5}{\beta^5_1} - \frac{\beta_0 f}{\beta_1}) \sim (1, 0,0)$ because of $\beta_0 f \sim 0$. The wave function is $\Psi= u_1 \psi^s_1+ u_2 \psi^s_2 +u_3 \psi^s_3= (\phi_1 +\phi_{2N})/\sqrt{2}$. The electrons are localized  at outermost layers of the ABCNLG  with $N=3$.
The  eigen state  $(u_1 ,u_2 ,u_3, u_4 )^T$ of the ABC-stacking quad-layer graphene is gained by solving the eigen-equation
\begin{eqnarray}
\left[
\begin{array}{cccc}
0 & \beta_0 f  &0   &0   \\
\beta_0 f &0 & \beta_1      &0    \\
0 & \beta_1 &0 &  \beta_0 f  \\
0 & 0 &\beta_0 f  & \beta_1   \\
\end{array}
\right]
\left[
\begin{array}{c}
u_1 \\
u_2 \\
u_3 \\
u_4 \\
\end{array}
\right]
=\lambda
\left[
\begin{array}{c}
u_1 \\
u_2 \\
u_3 \\
u_4 \\
\end{array}
\right].
\label{eigen eq-4}
\end{eqnarray}
The  eigen state $|\textbf{\emph{u}}_{\xi}\rangle^T $ is $(u_1 ,u_2 ,u_3, u_4 )=(1,(\frac{\beta_0 f}{\beta_1})^3,  (\frac{\beta_0 f}{\beta_1})^7 -\frac{\beta_0 f}{\beta_1}, -(\frac{\beta_0 f}{\beta_1})^{10} + (\frac{\beta_0 f}{\beta_1})^4  + (\frac{\beta_0 f}{\beta_1})^2 )\sim (1, 0,0,0)$. The wave function is $\Psi= u_1 \psi^s_1+ u_2 \psi^s_2 +u_3 \psi^s_3+u_4 \psi^s_4= (\phi_1 +\phi_{2N})/\sqrt{2}$, i. e., the electrons are presented in the outermost layers.

\section{Density of state of an ABCNLG }

Density of state (DOS), which reveals the main features of  the electronic structures,  is defined as
\begin{eqnarray}
D(\omega)=\frac{2}{\pi}\sum_{j=1}^{j=2N}\int\limits_{1st
BZ}\frac{d^2\mathbf{k}}{(2\pi)^2}{\frac{\Gamma}{(\lambda_j(\mathbf{k_x, k_y})-\omega)^2-\Gamma^2}}.
\end{eqnarray}
$\Gamma$ (=0.001 $\beta_0$) is the phenomenological broadening parameter.  The summation $\Sigma$ runs over all the subband index $j$,  starting at  $j=1$ and ending at $j=2N$. The $2N$ subbands are divided into two groups, symmetrical  and asymmetrical  groups. The summation is rearranged as $\displaystyle \sum_{j=1}^{2N}= \sum_{j_s=1}^{N} + \sum_{j_a=1}^{N}$ and  DOS $D(\omega)$ is decomposed into
two parts $D_s(\omega\,)$ and $D_a(\omega\,)$. That is:
\begin{eqnarray}
D(\omega\,)=D_s(\omega\,)+D_a(\omega\,),
\end{eqnarray}
where $D_s(\omega\,)= -D_a(\omega\,)$  because of
$\lambda_s= -\lambda_a$.

DOS of $N=3$, 4, 5, and 6 ABCNLGs are presented by the black curves  in Figs. 4(a)-4(d). The low energy DOS of $N=3$, 4, 5, and 6 ABCNLGs are exhibited in Fig. 5. The special features in the DOS, including sharp peak,  square-root divergences, logarithmic singularity, and  discontinuity are immediately associated with the structures of the energy dispersions [Figs. 2 and 3]. The flat bands, as is  shown in Fig. 3, give rise to a sharp peak located at $E_F=0$ [Fig. 5]. Energy bands near $\pm \beta_1~(\pm 0.36$ eV), are  presented as concave-upward (concave-downward) parabolic bands, leading to square-root divergences. DOS has $N$ logarithmic peaks near $\omega \sim \pm\beta_0$ (3.16 eV) [insets in the right panel of figure 4], resulting from the separated saddle points near the $M$ point in Fig. 2.  The maximum (or minimum) of the  energy dispersions occurring at the point $\Gamma$ [Fig. 2] brings about the discontinuity  at $\omega \sim \pm 3 \beta_0 $.

The red (green) curves in Figs 4 and 5 are  DOS $D_s(\omega\,)$
($D_a(\omega)$)  related to the energy bands of symmetrical (asymmetrical) group, the eigenenergy of Hamiltonian  $\mathbb{H}_{s}$ ($\mathbb{H}_{a}$). The calculated results illustrate that   $D_s(\omega)\ne D_s(-\omega)$ and $D_a(\omega)\ne D_a(-\omega)$, and
the sub-DOS $D_s(\omega)$ ($D_a(\omega)$)
is not symmetry about $E_F=0$. The red curves are symmetrical to the  green ones about  $E_F=0$, i. e.,  $D_s(\omega)= D_a(-\omega)$. The low energy DOS, presented in Fig. 5, exhibits that whether the layer number $N$ is odd or even has a great effect on the detail structures of DOS.

\section {CONCLUSIONS}
The electronic properties of the ABC-stacking  $N$-layer graphenes are explored through the tight-binding method, based on  the minimal model, including only the main inter- and intra-layer interactions. An ABCNLG can be modeled as a linear atomic chain with $2N$ atoms and described by a complex $2N \times 2N$ tridiagonal matrix $H_0$, which is further reduced into two real $N \times N$ tridiagonal matrices $\mathbb{H}_{s}$ and $\mathbb{H}_{a}$, i. e., $H_0=\mathbb{H}_{s} \oplus \mathbb{H}_{a} $ through a unitary transformation. Most importantly, the two matrices are shown to have the relation $\mathbb{H}_{a}=-\mathbb{H}_{s}$.  The eigenenergies of $\mathbb{H}_{s}$  and those of $\mathbb{H}_{s}$ have the relation $\lambda_{a}=-\lambda_{s}$. The transformation of   a complex $2N \times 2N$ tridiagonal matrix $H_0$ into a real $N \times N$ tridiagonal matrix allows us to efficiently explore  the full band structure, low energy dispersions and DOS. Above all,  our analytical model can be utilized to study the eigenvalue problem of linear chain model and the tridiagonal Toeplitz matrix.

\section*{Acknowledgement}
The authors gratefully acknowledge the support of the Ministry of Science and Technology of Taiwan, under Grants No. MOST105-2112-M-165-001-MY3.\\

\newpage
{\Large\bf Figure Captions}\\
\begin{itemize}
\item[FIG. 1.] (a)  exhibits the geometric structure, the intralayer interaction, and the interlayer interaction of  ABC-stacking trilayer graphene. The hexagonal ring is the first Brillouin zone.  (b) The ABC-stacking trilayer graphene is modeled by  a six-atom  chain.  Along the chain,   the  nearest neighbor interactions $\beta_0 f$ and $\beta_1$   are labeled.   The  six-atom  chain is cut into two  3-atom  chains.  The site energy of  the surface atom, marked by red circle,  is  $\beta_0 f$ (or  $-\beta_0 f$).  (c)  The same plot as (b) but for the ABC-stacking quad-layer graphene.

\item[FIG. 2.] (a)  The full  energy band of the ABC-stacking trilayer graphene.  (b) The same plot as (a) but for the ABC-stacking quad-layer graphene.

 \item[FIG. 3.] (a)-(c)  present the  low energy dispersions of the ABC-stacking multilayer graphenes with  layer numbers $N=3$, 5,  and 7.  The  low energy dispersions of the ABC-stacking multilayer graphenes  $N=4$, 6,  and 8 are shown in (d)-(f).  The red (green) curve illustrates  the energy spectrum of  the reduced Hamiltonian matrix $\mathbb{H}_{s}$ ($\mathbb{H}_{a}$).

 \item[FIG. 4.] DOS of the ABC-stacking multilayer graphenes with  layer numbers $N=3$,  4, 5,  and 6.   The red (green) curve is DOS related to $D_{s}(\omega)$ ($D_{a}(\omega)$).

  \item[FIG. 5.](a)-(d)  present the  low energy DOS of  ABC-stacking multilayer graphenes with  layer numbers $N=3$, 5,  7 and 15.  The  low energy DOS of ABC-stacking multilayer graphenes  $N=4$, 6,  8 and 16 are shown in (e)-(h). The sub-DOS $D_{s}(\omega)$ ($D_{a}(\omega)$) is marked by the red (green) curve .
 \end{itemize}

 \end{document}